\documentclass[sigconf, screen]{acmart}

\usepackage{enumitem}
\usepackage{listings}
\usepackage{wrapfig}
\usepackage{subcaption}
\usepackage{bbding}

\AtBeginDocument{%
  }

\copyrightyear{2023} 
\acmYear{2023} 
\setcopyright{acmlicensed}
\acmConference[CHI '23]{Proceedings of the 2023 CHI Conference on Human Factors in Computing Systems}{April 23--28, 2023}{Hamburg, Germany}
\acmBooktitle{Proceedings of the 2023 CHI Conference on Human Factors in Computing Systems (CHI '23), April 23--28, 2023, Hamburg, Germany}
\acmPrice{15.00}
\acmDOI{10.1145/3544548.3581452}
\acmISBN{978-1-4503-9421-5/23/04}

\acmSubmissionID{4741}

\begin{document}

\title{NetworkNarratives: Data Tours for Visual Network Exploration and Analysis}

\author{Wenchao Li}
\orcid{0000-0003-2605-7331}
\affiliation{%
  \institution{The Hong Kong University of Science and Technology}
  \city{Hong Kong SAR}
  \country{China}}
\email{wlibs@connect.ust.hk}

\author{Sarah Sch{\"o}ttler}
\orcid{0000-0002-4898-2619}
\affiliation{%
  \institution{University of Edinburgh}
  \city{Edinburgh}
  \country{United Kingdom}}
\email{sarah.schoettler@ed.ac.uk}

\author{James Scott-Brown}
\orcid{0000-0001-5642-8346}
\affiliation{%
  \institution{University of Edinburgh}
  \city{Edinburgh}
  \country{United Kingdom}}
\email{james@jamesscottbrown.com}

\author{Yun Wang}
\orcid{0000-0003-0468-4043}
\affiliation{%
  \institution{Microsoft Research Asia}
  \city{Beijing}
  \country{China}}
\email{wangyun@microsoft.com}

\author{Siming Chen}
\orcid{0000-0002-2690-3588}
\affiliation{%
  \institution{Fudan University}
  \city{Shanghai}
  \country{China}}
\email{siming.chen@fudan.edu.cn}

\author{Huamin Qu}
\orcid{0000-0002-3344-9694}
\affiliation{%
  \institution{The Hong Kong University of Science and Technology}
  \city{Hong Kong SAR}
  \country{China}}
\email{huamin@cse.ust.hk}

\author{Benjamin Bach}
\orcid{0000-0002-9201-7744}
\affiliation{%
  \institution{University of Edinburgh}
  \city{Edinburgh}
  \country{United Kingdom}}
\email{bbach@inf.ed.ac.uk}

\renewcommand{\shortauthors}{Li et al.}

\newcommand{\rev}[1]{\textcolor{purple}{#1}}
\newcommand{\reviewer}[1]{\textcolor{orange}{[\textbf{Reviewer:} #1]}}
\newcommand{\ben}[1]{\textcolor{red}{
#1}}
\newcommand{\yun}[1]{\textcolor{violet}{\textbf{yun:} #1}}
\newcommand{\sarah}[1]{\textcolor{magenta}{\textbf{sarah:} #1}}
\newcommand{\wenchao}[1]{\textcolor{cyan}{\textbf{wenchao:} #1}}
\definecolor{cobalt}{rgb}{0.0, 0.28, 0.67}
\newcommand{\james}[1]{\textcolor{cobalt}{\textbf{james:} #1}}


\newcommand{\numtours}{{10}}
\newcommand{\numfacts}{{102}}
\newcommand{\numfactsfromlit}{{102}}
\newcommand{\numparticipants}{14}
\newcommand{\tool}{NetworkNarratives}
\newcommand{\cguided}{\textsc{Tours}}
\newcommand{\cfreeform}{\textsc{Free-form}}
\newcommand{\etal}{~et~al.}
\sloppy

\newcommand{\says}[2]{\textit{``#1''} {\small[#2]}}
\newcommand{\tom}{\textit{Arch}}
\newcommand{\chris}{\textit{Hist1}}
\newcommand{\markw}{\textit{Soc1}}
\newcommand{\chamil}{\textit{Soc2}}
\newcommand{\lisa}{\textit{Health}}
\newcommand{\pyifan}{\textit{Hist2}}
\newcommand{\pjiangz}{\textit{Dev1}}
\newcommand{\pweic}{\textit{Dev2}}

\hyphenation{op-tical net-works semi-conduc-tor}

\newcommand{\formatgoal}[1]{\textsc{#1}}
\newcommand{\glearn}[0]{\formatgoal{Learn}}
\newcommand{\greduce}[0]{\formatgoal{Reduce}}
\newcommand{\gserendipidy}[0]{\formatgoal{Surprise}}
\newcommand{\grepeat}[0]{\formatgoal{Repeat}}
\newcommand{\gbalance}[0]{\formatgoal{Balance}}
\newcommand{\gtransparency}[0]{\formatgoal{Transparency}}

\begin{abstract}
This paper introduces semi-automatic data tours to aid the exploration of complex networks. Exploring networks requires significant effort and expertise and can be time-consuming and challenging. Distinct from guidance and recommender systems for visual analytics, we provide a set of \emph{goal-oriented} tours for network overview, ego-network analysis, community exploration, and other tasks. Based on interviews with five network analysts, we developed a user interface (NetworkNarratives) and 10 example tours. The interface allows analysts to navigate an interactive slideshow featuring facts about the network using visualizations and textual annotations. On each slide, an analyst can freely explore the network and specify nodes, links, or subgraphs as seed elements for follow-up tours. Two studies, comprising eight expert and 14 novice analysts, show that data tours reduce exploration effort, support learning about network exploration, and can aid the dissemination of analysis results. NetworkNarratives is available online, together with detailed illustrations for each tour. 
\end{abstract}



\maketitle


\section{Introduction}
\label{sec:intro}

Large, dense, and multivariate (potentially including node types or geographic locations, and link types, weights, times, or directions) relational datasets (networks) pose challenges to exploratory data analysis. Visualization interfaces for interrogating and exploring networks have grown increasingly sophisticated in order to support the richness of potential questions~\cite{lee_task_2006}. Tools such as Gephi~\cite{bastian2009gephi}, Palladio~\cite{palladio}, TempoVis~\cite{Ahn2011}, or Visone~\cite{baur2001visone}, provide different visual encodings, interaction, analysis metrics, and multiple visualization types, such as adjacency matrices, timelines, and maps that are sometimes presented as multiple coordinated views~\cite{bach2015networkcube}. 

Although powerful when used by an experienced analyst, feature-rich user interfaces present challenges for novice analysts who are required to learn possible interactions, understand the aim, perform interactions, and keep track of everything. 
Furthermore, significant time can be taken up by repeating steps, applying them to different datasets, keeping track of one's exploration, employing layout and exploration strategies, and undoing interactions in case of mistakes. In such free-form exploration interfaces, analysts can become lost or overwhelmed~\cite{yoghourdjian2020scalability} or make analysis errors, such as succumbing to the \textit{drill-down fallacy}~\cite{lee2019avoiding}. For a novice analyst, a rich set of tool features can result in a steep learning curve that requires cognitive effort to understand each feature and its affordances and effects~\cite{boy2015suggested}. As highlighted in a recent study~\cite{alkadi2022understanding}, the open-ended nature of exploration can be overwhelming to novice analysts who may not know what information can be gleaned from a network and what questions can be answered with network visualization. Therefore, creating effective exploration strategies and learning about network exploration without appropriate training or experience are challenging. 

\begin{figure*}
    \centering
    \includegraphics[width=\textwidth]{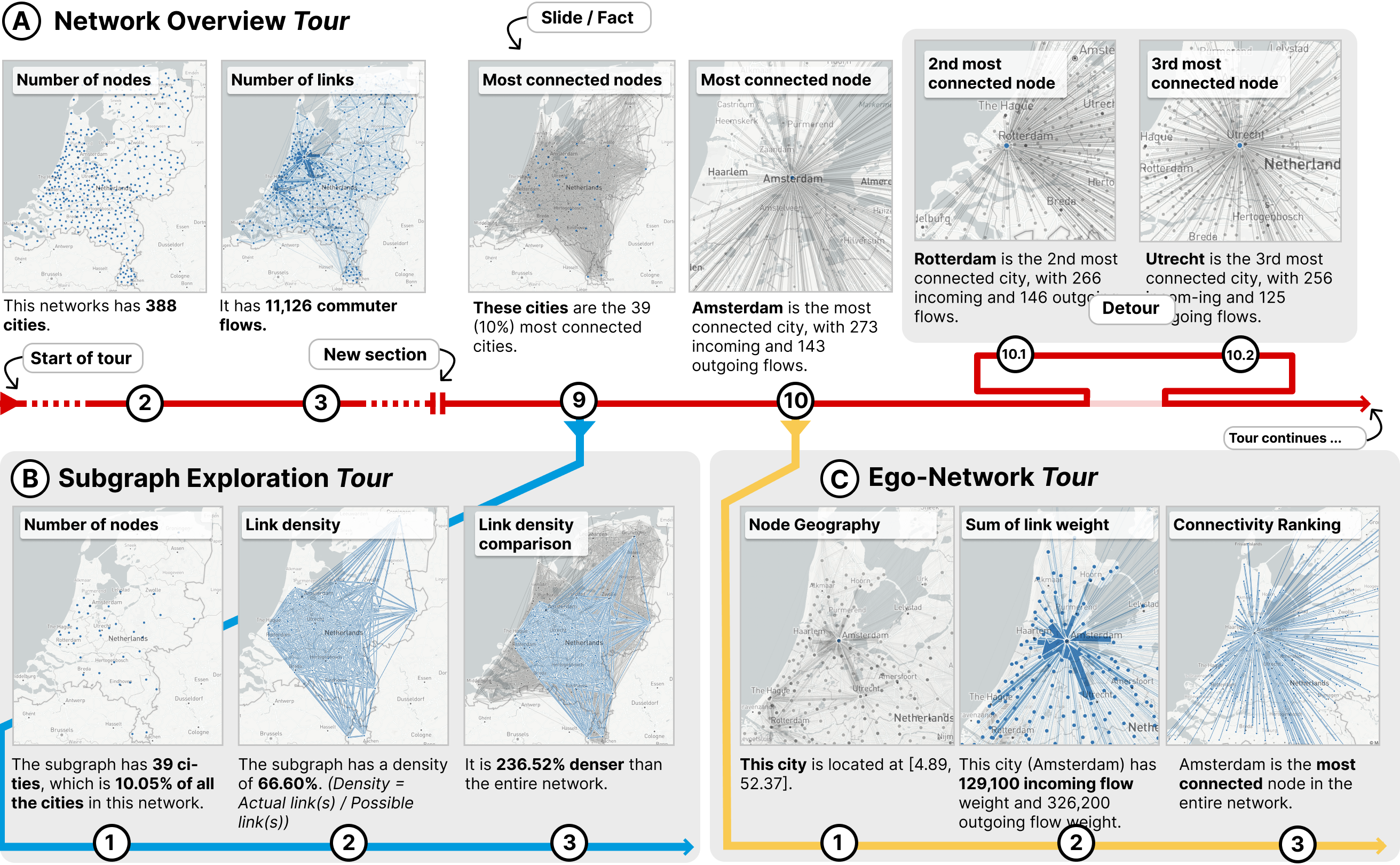}
    \caption{Conceptual illustration of data tours in \tool. An initial data tour of Network Overview (A, red line) showing facts (on slides) about the network. Each slide has a title and textual description. Numbers in circles indicate the number of facts in the tour. At times, e.g., when discussing a node or subgraph, a user can pivot to related tours (B, C) about that specific node or subgraph. Detours (D) include additional slides on demand for any given tour.
    }
    \Description{Figure 1 shows example data tours and how tours can be switched. }
    \label{fig:tours-concept}
\end{figure*}

In this work, we explore the idea of semi-automatic data tours to aid in network exploration. The idea of data tours goes back to Asimov's Grand Tour for multivariate data~\cite{Asimov1985} and has been recently described in a theoretic framework~\cite{mehta2017datatours}. 
We obtain further inspiration from ideas on guidance (e.g., as implemented in tools such as SocialAction~\cite{SocialAction} and Small Multiples~\cite{van2013small}), recommender systems~\cite{kim_graphscape_2017}, and data-driven storytelling (notably graph comics~\cite{bach2016telling} and interactive slideshows~\cite{segel_narrative_2010}). In our case, a \textit{data tour} walks an analyst through their network, similar to viewing a slideshow presentation created by another analyst (\autoref{fig:tours-concept}). Each slide in a tour (e.g., panels above the red line representing the \textit{Network Overview} tour) shows a specific piece of information about the network as explained in the caption. Our data tours aim to lower the barrier for novice analysts to learn interaction and exploration strategies and provide quick overviews of unknown datasets to expert analysts. 
Our data tours are defined by three main characteristics:

\textbf{First}, unlike existing recommender systems~(e.g., \cite{wang_datashot_2020,shi_calliope_2021}), our approach is \textit{goal-driven}, i.e., a data tour in our case is best thought of as a template representing an exploration strategy or story. We loosely describe an exploration or analysis strategy as \textit{``a set of information in a purposeful order, selected with the goal of providing insights into a dataset.''} For example, an analyst selects a subgraph and learns about the number of its nodes and density, as well as the most connected nodes and relations to the rest of the network (\autoref{fig:tours-concept}). To the best of our knowledge, this paper is the first to describe a practical approach to data tours in the domain of network analysis. 

\textbf{Second}, tours are primarily \textit{sequential}, allowing an analyst to easily flip through facts at their own pace while reducing the cognitive load imposed by decision-making about navigation and filtering. At any point, an analyst can take control, freely explore the network, and embark on a new tour or detours by inserting facts from \tool' recommender engine (\autoref{fig:tours-concept}(D)).

\textbf{Finally}, we implement \numtours{} complementary example data tours that are \textit{linked} so an analyst can pivot between them to change the focus of his exploration. For example, when a slide focuses on a specific node, the user is offered the opportunity to start a tour on the ego-network exploration of that node (\autoref{fig:tours-concept}(B)+(C)). Although not all presented facts in our data tours might be of interest to an analyst, we agree with Tukey regarding the importance of \textit{``notice what we never expected to see''}~\cite[Preface]{tukey1977exploratory}, \textit{``[to find] nothing, is a definite step forward''}~\cite[Preface]{tukey1977exploratory}, and to \textit{``give users a chance to think of initial questions [to help] them get started}''~\cite{north2006toward,toms2002information}.

Our concept of data tours is defined with \textbf{six design goals} in mind (\autoref{sec:designgoals}). Our \numtours{} individual tours are designed in collaboration with five network analysis experts with backgrounds in social science, history, epidemiology, and archaeology. We have implemented these tours in our user interface and recommender system \textit{NetworkNarratives}, which allows users to choose tours, navigate tours, and freely explore the network.
Qualitative feedback is obtained from eight network analysis experts who explored their own data with NetworkNarratives. In addition, a comparative study with \numparticipants{} novice analysts that suggests that our tours save time during the exploration process and that a goal-driven approach can make the exploration more accessible (\autoref{sec:evaluation}): tours provide a simple set of entry points and allow analysts to choose from a well-defined set of tours, each representing a specific analysis goal. 

In summary, our contributions are as follows: 
\begin{enumerate}[noitemsep,leftmargin=*]
    \item the concept of data-driven data tours for network analysis;
    \item \numtours{} extensible fully-implemented data tours (\autoref{sec:example_data_tours}), including \numfacts{} individual facts (\autoref{sec:characteristics}) for multivariate, temporal, and geographic networks;
    \item \textit{\tool}, an interactive user interface to experience data tours (\autoref{sec:ui_demo}), which is publicly available and can be used as either a standalone application or an extension to \textit{The Vistorian}~\cite{alkadi2022understanding} (documentation and video demos are also available online: \url{https://networknarratives.github.io});
    \item 
    two studies with 8 network analysis experts and 14 novice analysts respectively, which evaluate the usefulness and future potential of data tours and the \tool~system (\autoref{sec:evaluation}).
\end{enumerate}

\section{Related Work}
\label{sec:related_work}

\subsection{Data Tours}

Data Tours were first introduced by Asimov, whose Grand Tour~\cite{Asimov1985} was a computer-generated overview of a multivariate dataset that transitioned between scatterplots showing different projections~\cite{tukey1977exploratory}. Such tours have been created for time-varying data~\cite{yu2010automatic} and multivariate data~\cite{healey2012interest}. Two recent approaches model individual facts about a dataset as nodes of a graph, with links between graphs as possible relations. In Graphscape~\cite{kim_graphscape_2017}, these nodes are connected to form tours (or stories). Similarly, Mehta\etal~\cite{mehta2017datatours} described a theoretical framework for a hierarchical structure of facts for data tours, including staged transitions between individual facts/nodes. 

In this work, we explore data tours for the exploration and analysis of networks. Closest to our data tours is the Systematic Yet Flexible (SYF) concept by Perer and Shneiderman~\cite{SYF,SYF2}. This approach combines the structured exploration of social networks through a predefined set of seven \textit{steps} (overview, rank nodes, rank edges, plot nodes, plot edges, find communities, edge types), each featuring a set of \textit{tasks}, such as different node degree rankings. Each step is supported by a visualization such as a node-link diagram or a ranking visualization. NetworkNarratives builds on this concept by diversifying the notion of a single data tour to multiple (\numtours) tours, each focusing on a specific goal such as exploring an ego-network, comparing two subgraphs, or exploring temporal and geographic networks. In addition, NetworkNarratives tries to minimize manual interaction and provides detours on demand. 

\subsection{Onboarding, Guidance, and Storytelling}

Onboarding and guidance are closely related concepts that aim to help users of (visualization) systems, tools, and techniques. Onboarding can include step-by-step wizards, guided tours, video-based tutorials, help centers or overlays~\cite{stoiber2019visualization,stoiber2022perspectives}, or cheatsheets~\cite{wang2020cheat}. Here, data tours can be seen as a form of onboarding. Instead of onboarding onto a specific tool, system or visualization technique, our system introduces (novice) analysts to questions about networks and methods for visual exploration. 

Like guidance systems~\cite{ceneda_review_2019,ceneda_characterizing_2017,collins_guidance_2018,sperrle2022lotse}, data tours guide a user \textit{while} they are using a system or tool, rather than providing general \textit{a priori} resources and knowledge similar to on-boarding. Systems and techniques for guidance can be categorized based on whether they merely help \textit{orient} a user through overview and aid in building a mental map, suggest options to choose from (\textit{directing}), or \textit{prescribe} views and analyses~\cite{ceneda_characterizing_2017} through storytelling~\cite{schulz2013towards}. Data tours prescribe information and views that can be easily followed by clicking through a pre-defined sequence of visualization views. 
Guidance for the exploration of networks~\cite{Plaisant2003, signposts, Jusufi2011} has been implemented by suggesting nodes (\textit{orienting}) that may be of interest to a user based on his current selection~\cite{crnovrsanin_visual_2011,hutchison_navigation_2013}. These systems focus on analytical and operational knowledge, i.e., understanding analysis methods and the respective tools and systems. We see our data tours as a first step toward \textit{methodological knowledge} for exploration and analysis to explain questions, goals, and specific steps, especially when the user lacks any goals or hypotheses to inform their actions~\cite{alkadi2022understanding}.

\textit{Provenance} systems are similar to \textit{guidance} systems in that both represent many different possible visualizations of a dataset. However, guidance systems deal with visual representations that a user could potentially look at in the future, and provenance systems (e.g., StoryFacets~\cite{StoryFacets}) are concerned with the visualizations that have already been viewed. Data tours could be created from a previous exploration and analysis history by making insights generic, e.g., by turning them into templates and re-applying them to different datasets. 

Data tours also draw inspiration from data-driven storytelling, specifically, graph comics~\cite{bach2016telling} and slideshows~\cite{segel_narrative_2010}. We follow the pattern of an interactive slideshow~\cite{segel_narrative_2010} that combines elements of author-driven storytelling (the slides and tours) and reader-driven storytelling (interactive exploration, selecting nodes and graph for further tours, and navigating tours). Data tours could be seen as paths through a network, and analysts are encouraged to create their own tours. However, we did not design \tool{} as a story authoring tool nor an automatic storytelling system. Actual data-driven storytelling requires information about the context and audience to be effective.

\subsection{Data Fact Recommendation} 
Automatic fact extraction has been used in guidance to review large datasets~\cite{Law2020_insights}, start the exploration loop~\cite{wills2010autovis}, support communication (e.g.,~\cite{wang_datashot_2020,wang_infonice_2018,yu2010automatic}), and avoid drill-down fallacies~\cite{lee2019avoiding}.
The problem of automatically visualizing facts from data can be subdivided into: (i) identifying individual data facts (often referred to as \textit{insights}); (ii) suggesting visualizations for each fact; and (iii) presenting facts and visualizations alongside each other. \tool{} implements some of these concepts. 

First, systems to extract data facts have been summarized by Law\etal~\cite{Law2020_insights}. Data facts can be selected using statistical techniques through a separate procedure to generate facts for each of a pre-determined list of fact types. Facts can then be ordered based on a ranking score that is intended to represent subjective relevance, importance, or degree of interest (e.g.,~\cite{shi_calliope_2021,wang_datashot_2020,harris2021insightcentric}). 
To the best of our knowledge, all these systems explicitly focus on tabular data, rather than network or relational data as explored in this work. The facts we consider include common network metrics as well as topological features. 

Second, a number of systems automatically suggest either a single visualization or a set of visualizations that most appropriately depict a dataset (e.g.,~\cite{mackinlay1986automating, van2013small, moritz_formalizing_2018, hu2019vizml, zhao2020chartseer}); these systems have been summarized elsewhere~\cite{Law2020_insights}. Visual encodings can be chosen based on either rules expressing guidelines using constraint satisfaction~\cite{moritz_formalizing_2018} or machine learning methods trained on examples of human choices~\cite{hu2019vizml, zhao2020chartseer}. 
The systems allow users to interactively browse chart recommendations~\cite{wongsuphasawat_voyager_2016} or receive them as notifications~\cite{cui2019datasite}.
In \tool{}, we initially limit ourselves to node-link diagrams, rather than automatically choosing from a range of visualization types. Other visualization techniques, such as adjacency matrices for dense graphs, are likely to require additional explanations (e.g.,~\cite{srinivasan2018augmenting,wang2020cheat,martinez2020data}) to enable new users to read them properly and obtain \textit{visual insights}~\cite{Law2020_insights}. 

Finally, facts and visualizations must be related and presented to the user, possibly in a form that enables interactive exploration and personalization. For example, researchers have designed interfaces that group insights into panels based on category~\cite{harris2021insightcentric,Demiralp2017_foresight},
display dashboards~\cite{key2012vizdeck}, or generate infographic-like fact sheets of visualizations featuring textual explanations~\cite{wang_datashot_2020}. 
To organize data facts into sequences, graph-based approaches create a similarity graph from all facts and consequently select a sequence of visualizations (a path through the graph) based on minimizing edge weight~\cite{yu2010automatic,kim_graphscape_2017}. Data facts can also be ordered using the complex technique of logic-oriented Monte Carlo tree search~\cite{shi_calliope_2021}.
An alternative approach is to apply the same type of visualization to all variables in a dataset and then select those that show the most interesting patterns: this is the route taken by scagnostics~\cite{wilkinson2008scagnostics,scagexplorer} and GRID~\cite{seo2005rank} for tabular data, and magnostics for adjacency matrices representing network data~\cite{behrisch2016magnostics}.

All these approaches are predominantly \textit{data-driven}, in that they start by analyzing the data and then ask the user to navigate and express their preferences. By design, these systems have only a loose notion of the inherent human factors underlying exploration, such as research questions and methodologies (whether formal or informal). Our approach is based on sequential high-level \textit{goals} that prescribe a set of facts and their specific sequence presented in an interactive slideshow~\cite{segel_narrative_2010}. Our templates are human-created to ensure that they target specific network analysis goals and present data facts in a meaningful sequence.

\section{Design Goals and Characteristics}
\label{sec:design_goals_characteristics}
This section describes the methodology of informing data tours and \tool{} (\autoref{sec:methodology}), followed by the design goals (\autoref{sec:designgoals}) and the resulting main characteristics of our approach (\autoref{sec:characteristics}). 

\subsection{Research Methodology}
\label{sec:methodology}
Data tours and the \tool{} system were established by a three-step process of literature review, domain expert interviews, and iterative prototyping and design. 

\textbf{1) We examined existing task taxonomies} to obtain a systematic overview of facts and information relevant to network exploration~\cite{yang_what-why_2019, lee_task_2006,Ahn2011,munznerVisualizationAnalysisDesign2014}. We found \numfactsfromlit~facts, and two of the authors formally categorized them by applying tags, such as \textit{centrality}, \textit{temporal}, and \textit{connectivity}. Although we originally aimed for a clear taxonomy of facts, we found that the resulting taxonomy was too ambiguous and not helpful for our work. The tags were ultimately used to search for facts when we created data tours (\autoref{sec:implementation}). 

\textbf{2) We reviewed written reports in papers and blog posts describing network analyses} and publications conducted by domain experts in different subject areas (e.g., \cite{abel_quantifying_2014,testimonials}). From these reports, we extracted metrics and insights and noted the order in which they were reported, yielding an initial collection of data tours to power \tool' recommender engine.

\textbf{3) We interviewed five domain experts in network analysis}. The individual interviews lasted one hour on average and focused on exploration and analysis goals, current workflow methods, and tools used. Four of the five analysts had a background in the Humanities and Social Sciences and analyze networks on a daily basis: \tom{} (Associate Prof.) analyzes historical transport networks, focusing on transportation costs (link weight) within the network; 
\chris{} (Prof.) explores historical social networks extracted from newspapers and letters; 
\lisa{} (Prof.) explores networks of collaborations between researchers and how they influence the policy advice that they provide about disease response; and \markw{} (Assistant Prof.) and \chamil{} (Assistant Prof.) investigate community structures in the network of interactions between users on social media sites, a task that often involves working with multiple link and node types (both \markw{} and \chamil{} have a strong quantitative background). Across all interviews, we identified tasks (e.g., understanding paths, exploring ego-networks) and exploration strategies (e.g., following a path along a geographic feature such as a river). Experts confirmed that existing free-form approaches were tedious to use, especially for repetitive steps and routines. 

\subsection{Design Goals}
\label{sec:designgoals}
The design goals were based on our conversations with analysts, the literature, and our own experience in working with network analysts over many case studies. 

\textbf{G1: \glearn---Introduce exploration strategies, goals, and concepts to novice analysts}---In our interviews, we saw that exploration is rarely entirely open, and is typically influenced by factors including high-level research goals and prior knowledge about the data, as well as personal analysis protocols and methodologies. For example, \markw{} and \chamil{} start their analysis by calculating overall network metrics, such as density, fragmentation, average path length, and average degree. They then focus not only on individual nodes, their centrality metrics, and ego-networks, but also on smaller communities within the network. \chris{} and \tom{} expressed minimal interest in general metrics and instead focus on the detailed exploration of geographic regions (\tom) or on individual nodes and their ego-networks, individual links and their attributes, and specific time slices of a dynamic network (\chris).

In an open-ended exploration, novice analysts might lack specific high-level \textbf{goals} when exploring a network~\cite{alkadi2022understanding}.
They might also be unfamiliar with network \textbf{concepts} required to ``decode'' information from networks (such as clusters and communities, node degrees, the shortest paths, and link weight). To achieve an exploration goal, analysts need \textbf{strategies} that they can apply through interactions and reading patterns from visualization (such as selecting nodes, calculating metrics, searching for elements with specific characteristics, and comparing graph elements~\cite{lee_task_2006}). Data tours can automate and demonstrate some of these concepts by exemplifying exploration for any given dataset. 

\textbf{G2: \greduce---Reduce cognitive and manual exploration effort}---All the analysts reported using and frequently switching between multiple tools to alternate between network visualizations (for topology and overview tasks) and metric calculation (for analysis tasks). These tools included \textit{igraph}~\cite{igraph} and \textit{tidygraph}~\cite{tidygraph} for calculating network analysis metrics and producing static visualizations, \textit{Visone}~\cite{baur2001visone} and \textit{Gephi}~\cite{bastian2009gephi} for interactive visualizations, GIS software for geographic data, and tools that they have written themselves. Switching tools interrupts work and requires piecing together information from different representations. For example, an analyst might use a script to find the node with the highest degree, and then switch to a network visualization tool to explore its context (\says{I use visualization for initial exploration, but to actually go into the analytics is difficult because it's messy}{\markw}). The analysts frequently export subnetworks as separate datasets for further analysis but report difficulties keeping track of these exported files. In addition, an exported subnetwork cannot easily be linked back to its context in the main network: \says{[these are] important needs, now that we have these very large databases}{\tom{}}.

Data tours can free an analyst from the majority of manual and cognitive labor required: 
a) interactions (e.g., selecting a time range, panning and zooming to find an element or visual pattern, or hovering over nodes to show their labels), 
b) decisions (e.g., what information to look at/what question to ask), 
c) visual search (e.g., finding the node with the highest degree, or finding the strongest or weakest link), and 
d) visual counting tasks (e.g., counting the degree of a node, or the number of unconnected nodes). 

\textbf{G3: \grepeat---Repeat routine explorations}---Data tours can easily be repeated and reapplied to different networks.
This feature can be useful when an analyst wants to repeat an analysis workflow with an updated dataset, apply the same exploration to subsets (e.g., connected components or clusters) of the complete dataset, or compare multiple versions of the same network (e.g., obtained by applying different filtering nodes and links). One participant reported \says{we ended up with tens of thousands of all sorts of networks}{\chris}.



\textbf{G4: \gbalance---Balance prescription and agency}---Data tours need to be inspiring, not restricting. Similar to guidelines, they should give direction and provide detailed steps into that direction. Analysts may change their priorities and shift their high-level goals: \says{the more you produce networks, the more you have a chance to really get lost; you don't know exactly what you have}{\chris}.

\textbf{G5: \gserendipidy---Support serendipitous discovery}---Despite being goal-oriented, an exploration process needs to remain open to serendipity and unexpected discoveries. Data tours should include a range of information (time, link-weight, and isolated nodes), especially when networks are large and multivariate: \says{with my current tool, I use random [strategies] and found it difficult to scale [exploration] at a reasonable size}{\markw}. 


\textbf{G6: \gtransparency---Keep information in tours and tour structure simple and transparent}---Our data tours follow regular templates scripted by a human author (\autoref{sec:implementation}). We want to avoid complex recommender models whose decisions might not be transparent to the analyst. Our tours are predictable in that they follow this human script, rather than a user model that tries to learn a user's intentions or can trap an analyst in a ``recommender bubble''. 

Tours for other datasets, audiences, and purposes might be designed with different goals in mind and implement different design decisions.

\subsection{Main Characteristics at a Glance}
\label{sec:characteristics}

The design goals \textbf{G1}--\textbf{G6} led us to the following design decisions. \autoref{fig:tours-concept} displays three example tours and their relationships, each one shown along a colored line: \textit{Network Overview} (red, top), \textit{Subgraph Exploration} (blue, left), and \textit{Ego-Network Analysis } (yellow, right). The Network Overview tour starts by explaining the number of nodes and links in the network, followed by the most connected nodes and the most connected node. In the following section, we explain the main tour concepts illustrated in \autoref{fig:tours-concept}.

\paragraph{\textbf{Goal-Oriented}}
Based on expert interviews and our own experience, we designed 10 complementary tours described in detail in \autoref{sec:example_data_tours}. In response to design goals \glearn{} and \greduce, each tour has a specific goal that is expressed by its title and abstract. For example, the goal of the \textit{Network Overview} tour (\autoref{fig:tours-concept}(A)) is to provide an overview of the most important facts about this network, such as the number of links and nodes, clusters, or link density. Goals can help inform an individual which tour to choose and why to embark on a specific tour. 

\paragraph{\textbf{Facts and Slides}}
Each tour is made of a sequence of \textit{facts}. A fact is a piece of information about the network such as the number of nodes and links, network density. Facts are expressed as templates with the respective values calculated from the dataset (e.g., \texttt{This network has N nodes}). In \tool, a fact is visually represented as a \textit{slide} (panels in \autoref{fig:tours-concept}) that shows a \break node-link visualization of that network, potentially highlighting \nobreak relevant nodes and links, and describing the fact in a caption. A \nobreak caption also explains unknown concepts, e.g., explaining \textit{link-density} alongside a link to a related web resource (\glearn). 
The number and sequence of facts in each tour are predetermined to serve a goal (\glearn). Depending on the data characteristics (e.g., temporal, geographic, link weights), irrelevant slides are removed from the tour by identifying the specific tags of topics covered in a fact.
Once a tour has been developed as a template (\autoref{sec:implementation}), it can be applied to many different datasets (\grepeat). Facts are automatically populated with metrics and visualizations computed from the provided data. 

\paragraph{\textbf{Sequence and Structure}}
To facilitate understanding of the tour and navigation (\gtransparency), slides (facts) for each tour are presented in a single, sequential order (\autoref{fig:tours-concept}). Slideshows are designed to provide cohesion among the facts and reflect the specific steps of an exploration strategy (\glearn). For example, a data tour might start with a high-level overview of basic network metrics and then explain one aspect in more detail while potentially linking to other tours. Alternatively, it might start with a single node and then zoom out to explore its ego-network and further to progressively introduce the surrounding nodes. A data tour could also follow a set of nodes on a path, highlighting connections between them (\autoref{sec:example_data_tours}). 
Facts about a similar theme are grouped into \textit{sections} to structure a tour into meaningful units and support fast navigation between sections (\greduce). For example, the Network Overview tour (\autoref{fig:tours-concept}) is subdivided into the sections \textit{Overview} (e.g., number of nodes and links, network density, and most connected nodes), \textit{Link information} (e.g., strongest link and weakest link), \textit{Nodes and centralities}, and \textit{Node clusters}. 

\paragraph{\textbf{Navigation}}
Navigation through the facts and sections of a tour is mainly linear (\greduce). At specific points, an analyst can pivot into a related tour. For example, the 4th slide in the red tour in \autoref{fig:tours-concept} mentions the most connected node in that network. An analyst can now decide to launch a tour on this node's ego-network, that is, launching the \textit{Ego-Network Analysis} tour (yellow). Meanwhile, the focus of \tool{} is not to be an open-ended exploration system. We provide freedom for user-driven exploration by allowing for limited personalization and ``detours'' from the set path of a data tour (\autoref{sec:ui_demo}, \gbalance). Detours are possible by inserting related facts into an ongoing tour, as suggested by \tool' recommender engine. For example, following the most connected node in the Network Overview tour, (4th panel in \autoref{fig:tours-concept}) the detour inserts two slides about the 2nd and 3rd most connected nodes in that (\autoref{fig:tours-concept}(D)). 


\section{Example Data Tours}
\label{sec:example_data_tours}
\tool{} currently implements 10 tours, whose goals and main facts are detailed as follows. The list is not meant to be exhaustive but show the richness of possible tours. The full list of all tours, including their facts, as well as some illustrations can be found online: \url{https://networknarratives.github.io/tours}. 

\begin{wrapfigure}{l}{.4cm}
    \vspace{-0.45cm}
    \includegraphics[width=1cm]{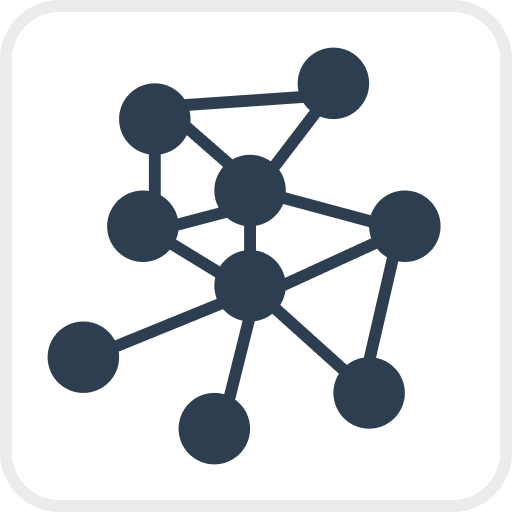}
    \vspace{-0.9cm}
    \end{wrapfigure}\noindent\textbf{Network Overview} describes an entire network. It starts with four introductory slides, covering the geographic extent (skipped for non-geographic networks), number of nodes and links, and the density. The second section focuses on links, showing the total and average link weight and the strongest and weakest links in the network. The third section provides some details about node centralities and the overall community structure (number of clusters).

\begin{wrapfigure}{l}{.4cm}
    \vspace{-0.45cm}
    \includegraphics[width=1cm]{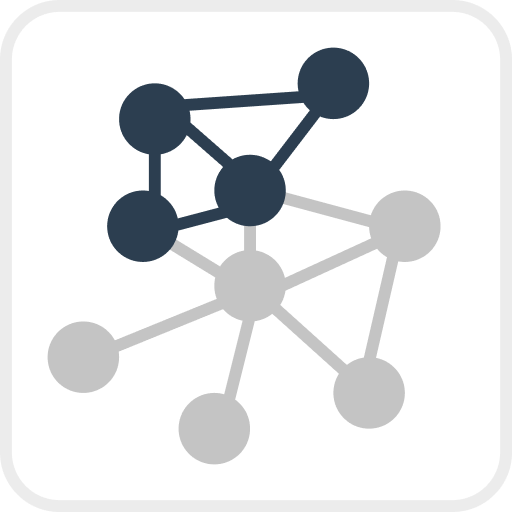}
    \vspace{-0.9cm}
    \end{wrapfigure}\noindent\textbf{Subgraph Overview} is similar to Network Overview, but focuses on a specific subgraph. It shows the subgraph's size and the percentage of the network's nodes. The tour also comprises important nodes such as the \textit{Most connected node} in the subgraph, \textit{Subgraph density}, and important links to the rest of the network. 

\begin{wrapfigure}{l}{.4cm}
    \vspace{-0.45cm}
    \includegraphics[width=1cm]{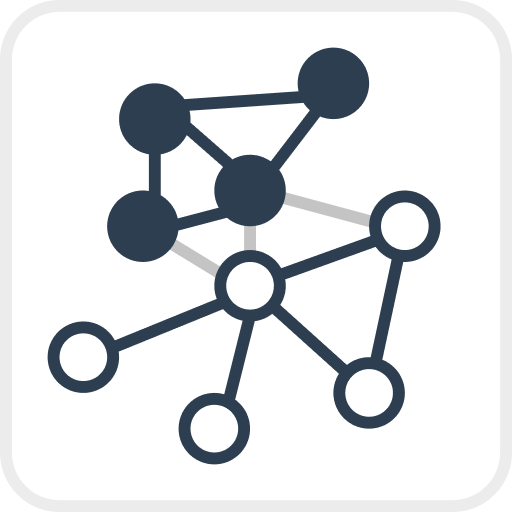}
    \vspace{-0.9cm}
    \end{wrapfigure}\noindent\textbf{Community Exploration} explores and compares clusters in the network and shows their sizes, connections, and important nodes. For example, the \textit{Most connected cluster} displays the cluster that has the most connections with the others. For community detection, we use the algorithm by \nobreak Newman~\cite{newman2004fast}. Advanced community detection algorithms can easily be included and used for comparison (e.g., $k$-means with different values for $k$) and shown on different slides (e.g., one slide for each value of $k$).
    
\begin{wrapfigure}{l}{.4cm}
    \vspace{-0.45cm}
    \includegraphics[width=1cm]{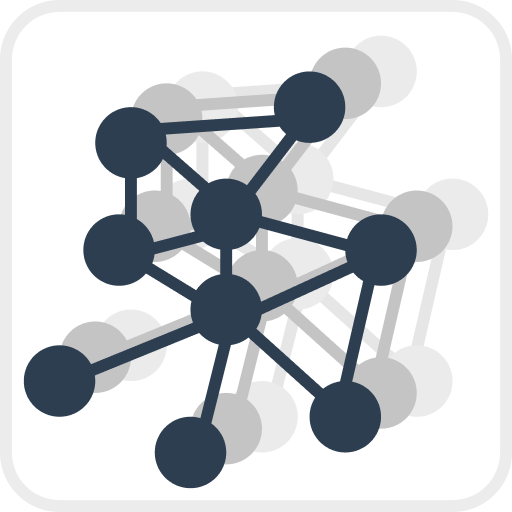}
    \vspace{-0.9cm}
    \end{wrapfigure}\noindent\textbf{Centrality Exploration} explores nodes based on different centrality measures (e.g., degree or betweenness). For example, we compute the \textit{Average degree centrality}, the node with the \textit{Highest betweenness centrality}, and the node with the \textit{Highest Closeness centrality}. Possible extensions include comparisons of several centrality measures. 


 \begin{wrapfigure}{l}{.4cm}
    \vspace{-0.45cm}
    \includegraphics[width=1cm]{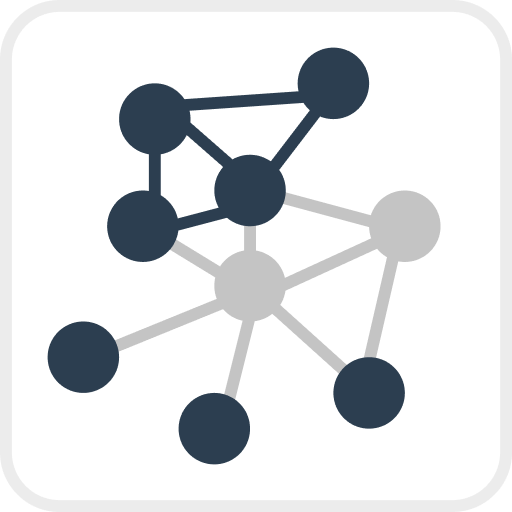}
    \vspace{-0.9cm}
    \end{wrapfigure}\noindent\textbf{Subgraph Comparison} compares two specified subsets of nodes and links (e.g., regions, subgraphs). Selecting this tour prompts the user to select two sets of nodes. The data tour first mentions the \textit{Number of nodes} and the \textit{Number of links} for each subgraph, then details important nodes such as the \textit{Most connected node} in each subgraph, and finally reveals links between the two subgraphs (e.g., \textit{Number of links}, \textit{Total link weight}, and \textit{Strongest link}).

 \begin{wrapfigure}{l}{.4cm}
    \vspace{-0.45cm}
    \includegraphics[width=1cm]{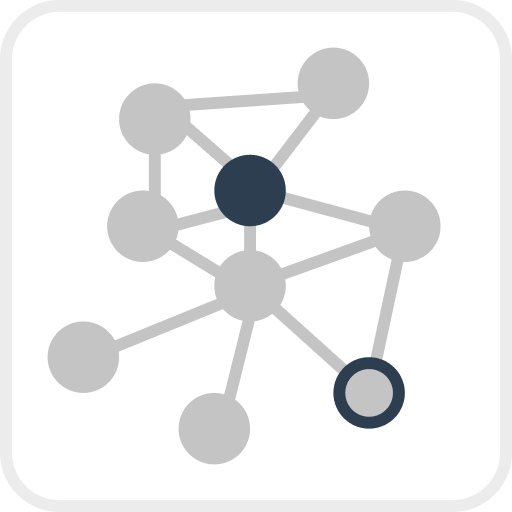}
    \vspace{-0.9cm}
    \end{wrapfigure}\noindent\textbf{Compare Two Nodes} shows the links between the two nodes, compares their connectivities and total link weights, and finally shows the common neighbors. For example, general statistics such as \textit{Connectivity ranking} and \textit{Total link weight} of the two nodes are compared. Neighboring nodes that connect both of the selected nodes are shown in the last slide (\textit{Common neighbors}) of the tour. 

 \begin{wrapfigure}{l}{.4cm}
    \vspace{-0.45cm}
    \includegraphics[width=1cm]{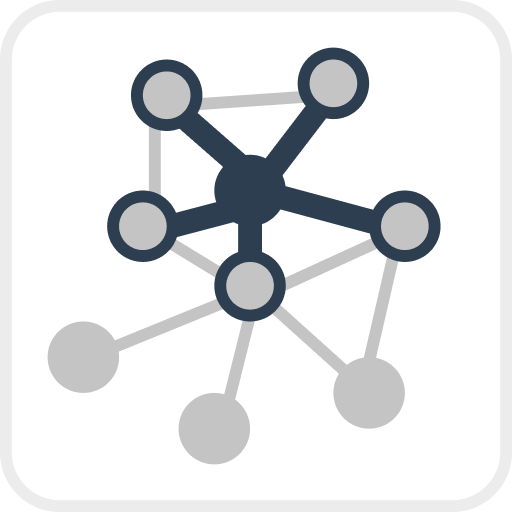}
    \vspace{-0.9cm}
    \end{wrapfigure}\noindent\textbf{Ego-Network} explores the network around a selected node and its neighbors. The data tour starts with the selected node and its position within the entire network. The tour then shows the node's direct neighborhood (nodes, links, strong connections), followed by their mutual connections, and finally its neighbors' neighbors. 

 \begin{wrapfigure}{l}{.4cm}
    \vspace{-0.45cm}
    \includegraphics[width=1cm]{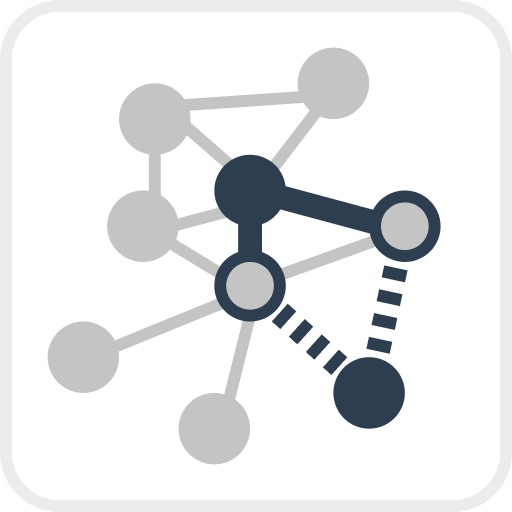}
    \vspace{-0.9cm}
    \end{wrapfigure}\noindent\textbf{Possible Paths} explores a set of possible paths between two selected nodes. The data tour reports the path length, combined weights along each path, and the minimum link weight within each path. This data tour is motivated by \tom{}'s desire to explore historical travel costs between cities. 

 \begin{wrapfigure}{l}{.4cm}
    \vspace{-0.45cm}
    \includegraphics[width=1cm]{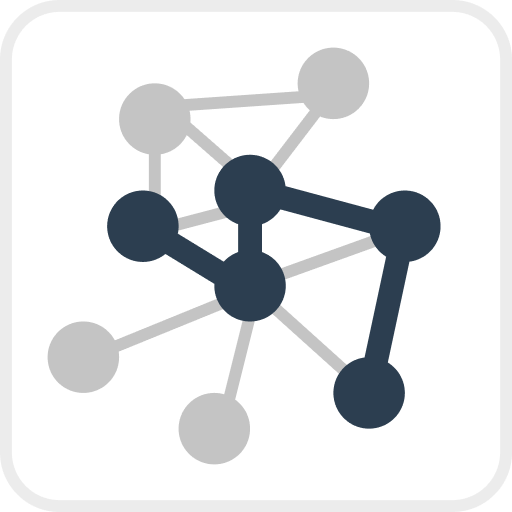}
    \vspace{-0.9cm}
    \end{wrapfigure}\noindent\textbf{Follow a Path} requires a selection of a set of connected nodes. The tour follows the path, explaining details about each node and its neighbors, and provides overall statistics of all the nodes in the path. This tour is motivated by \tom{}'s interest in nodes along geographic features, such as rivers, main roads, or political boundaries.

 \begin{wrapfigure}{l}{.4cm}
    \vspace{-0.45cm}
    \includegraphics[width=1cm]{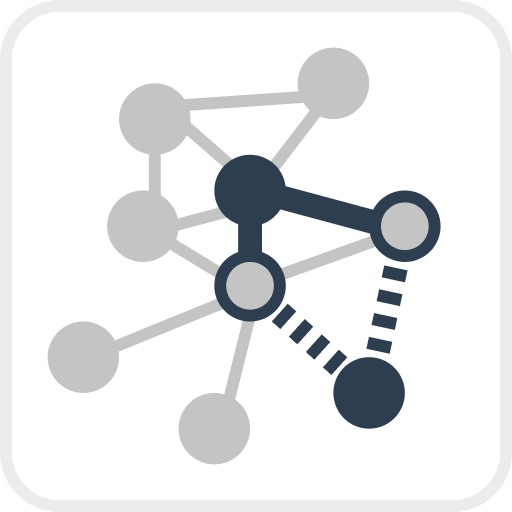}
    \vspace{-0.9cm}
    \end{wrapfigure}\noindent\textbf{Temporal Exploration} starts with an overview, showing the basic statistics about the network. It then demonstrates the connectivity evolution of the network based on temporal attributes. The data tour  ends with the comparison of \textit{Network density} over different time periods.






\section{NetworkNarratives User Interface}
\label{sec:ui_demo}
To explore the potential of our guided data tours and allow for real-world use, we built a web-based prototype system called \textit{\tool{}}. \tool{} can visualize and create tours of geographic and temporal networks that may include multiple link types and link weights. Networks are shown on a map if geographic information is given. Otherwise, they are rendered by using a force-directed layout. 

The user interface (\autoref{fig:poc}) of \tool{} consists of six panels: 
(a) data upload and selection; 
(b) a list of available data tours; 
(c) a detailed outline of the content of the currently selected data tour; 
(d) the network visualization with (e) captions; and (f) a navigation panel. 
Below, we explain how a user applies \tool{} to explore a dataset of commuter movements in the Netherlands~\cite{commuting_data}. 

\begin{figure*}[t]\setcounter{figure}{1}
    \centering
    \includegraphics[width=0.98\textwidth]{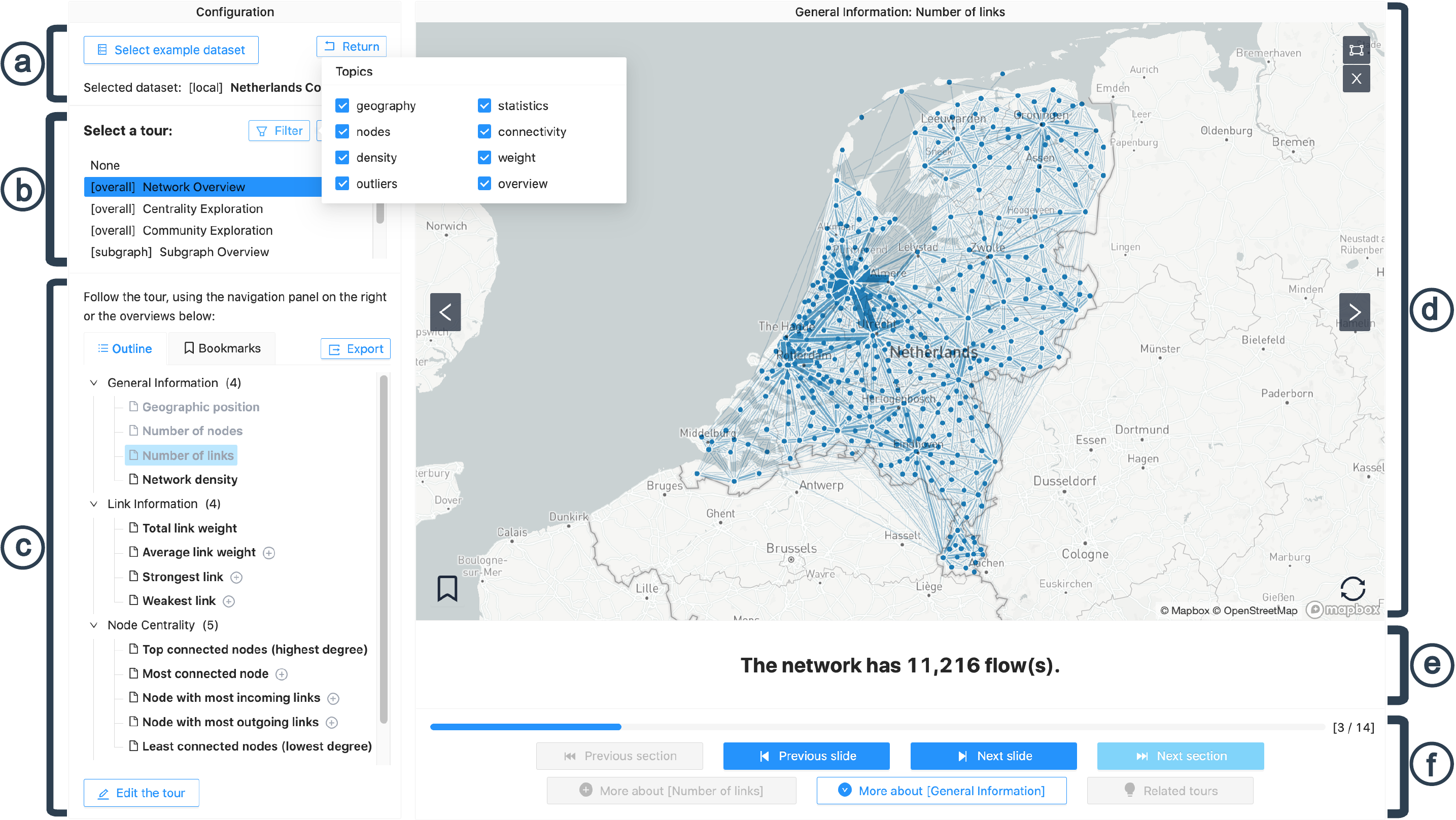}
    \caption{
    User interface of \tool{}. The left sidebar contains the data selection panel (a), data tour selection panel (b), and tour overview and starred slides panel (c). Each slide consists of an interactive visualization (d) accompanied by a textual description (e). The user can move between slides by clicking on the arrow buttons in the visualization panel (d), using the buttons in the navigation panel (f), or clicking on the slide names in the outline panel (c).
    }
    \Description{Figure 2. Fully described in the text. }
    \label{fig:poc}
    \vspace*{-6pt}
\end{figure*}

\subsection{Importing data and defining  terminology}
After import, the analyst is prompted to specify domain-specific terminology for nodes, links, link-weight, and subgraphs. To \break illustrate, in the commuter network example, they might refer to nodes as \texttt{cities}, links as \texttt{flows}, and link weight as \texttt{number of commuters}. This feature was added in response to one of our analysts and aims to help relate explained facts easily to the \nobreak domain~\cite{boy2014principled}. 

\subsection{Choosing and scoping a data tour}
A user can then choose any data tour from the selection panel (\autoref{fig:poc}(B)). Hovering the cursor over a data tour's name displays a tooltip containing a short description. This approach simplifies browsing and selecting tours of interest (\glearn, \gserendipidy, \grepeat). Alternatively, the analyst can select nodes and subgraphs in the visualization view by clicking on nodes or using a lasso interaction and subsequently choosing a data tour exploring the selected subgraph or node (e.g., \textit{Ego-Network Exploration}, \textit{Subgraph Overview}, and \textit{Subgraph Comparison}). 

\subsection{Starting a data tour}

Once a tour is selected from the tour panel in \autoref{fig:poc}(A), its structure becomes visible in the outline panel (\autoref{fig:poc}(C)) showing a tour's sections and facts in a tree-view. The visualization view shows a popup with the title and description of the tour and prompts the user to start the tour. The example in \autoref{fig:tours-concept} starts with the \textit{Network Overview} tour.

Clicking the ``\textit{Click to start}'' button in the popup loads the first \textit{slide} in visualization view. Each slide refers to a single fact in the network. The first row in \autoref{fig:tours-concept} shows selected screenshots from the Network Overview tour. A slide includes \textit{i)} a title representative of the fact shown on that respective slide (e.g., ``Number of Nodes''), \textit{ii)} an interactive visualization of the network (\autoref{fig:poc}(D)), and \textit{iii)} a caption stating the corresponding fact (e.g.,\textit{ ``This network has 11,216 links.''}). Any specific nodes or links mentioned by a fact are highlighted in the visualization. 

Whenever a fact mentions a term that might be unfamiliar to the novice analyst or that might require additional explanation (e.g., link density), \tool{} displays a hyperlink for a popup window. In the case of link density, the popup explains the formula used to calculate link density in \tool{} (\gtransparency, \glearn). 

\subsection{Navigating a data tour}
An analyst can navigate to the next slide in the data tour by using the \includegraphics[height=9pt]{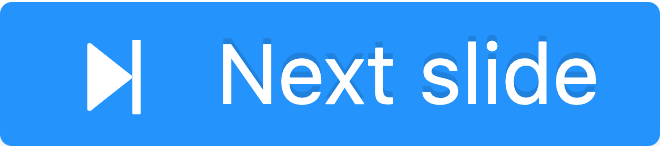} button. 
A popup over the visualization indicates when the analyst has reached the second section of the tour. They can continue stepping through the slides by using the \includegraphics[height=9pt]{images/btn-next-slide} button, or choose to skip that section and press the \includegraphics[height=9pt]{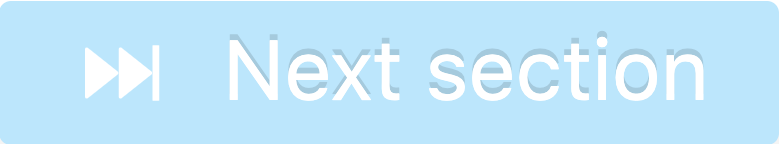} button. Alternatively, the analyst can return or jump directly to any specific slide in a section by clicking on its title in the outline view (\autoref{fig:poc}(C)). 

In each slide, the network visualization is interactive to allow for exploration and the specification of nodes and links for potential tours (\gtransparency). Within a visualization, an analyst can pan, zoom, and hover to highlight the connections of a specific node. 

\subsection{Detours}

\begin{figure}[ht]
  \centering
  \includegraphics[width=0.48\textwidth]{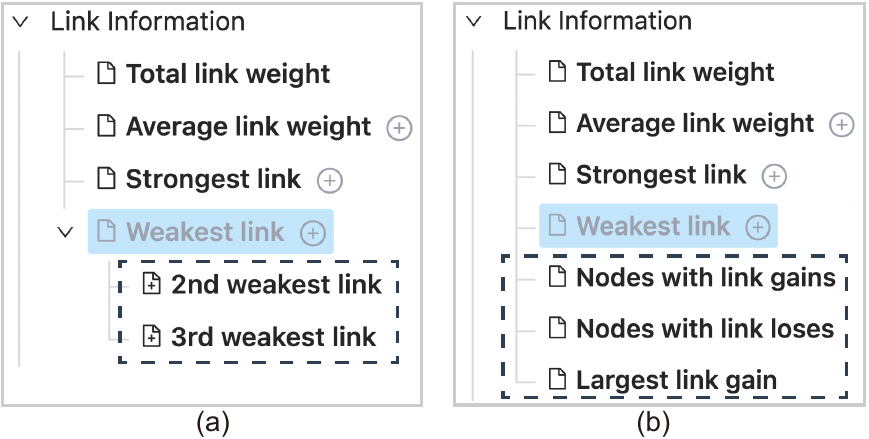}
  \caption{Slides and sections can be extended to show additional related slides. This example shows the results of extending (a) the \textit{Weakest link} \textbf{slide} and (b) the \textit{Link Information} \textbf{section}. In both images, the \textit{Weakest link} slide is selected for display. }
  \Description{Figure 3. Fully described in the text. }
    \label{fig:more_slide}
\end{figure}


A detour inserts additional facts into a tour. To include a detour, an analyst can click the \includegraphics[height=9pt]{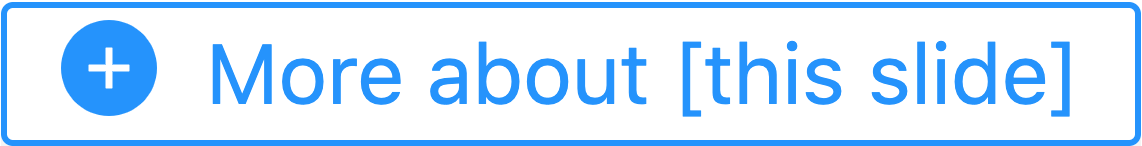} button in the navigation panel. For example, if they are viewing the slide about the strongest link, then two new slides for the second and third strongest link are inserted (\autoref{fig:more_slide}(a)). Clicking the \includegraphics[height=9pt]{images/btn-more-slide} button again appends additional relevant slides from \tool' fact library that are recommended by the system (see \autoref{sec:implementation} for details). Likewise, detours can be included on a section level, i.e., to extend the current section. Clicking the \includegraphics[height=9pt]{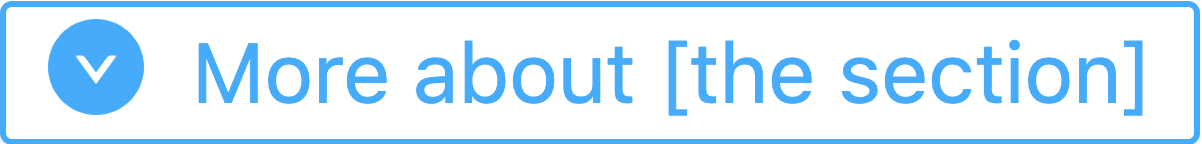} button inserts slides with relevant facts about that section that are recommended by \tool{} (\autoref{fig:more_slide}(b)). 

Finally, the analyst can filter slides about facts that are not of interest. By clicking the ``\textit{filter}'' button in the data tour selection panel, the analyst can uncheck any irrelevant aspects (e.g., \textit{time} and \textit{link weight}). Internally, these topics are stored in the form of \textit{tags} associated with the facts and used to suggest related facts (\autoref{sec:implementation}).

\subsection{Pivoting to related data tours}
Pivoting is possible for any slide discussing specific nodes or subgraphs. In these cases, \tool{} displays a small popup that suggests to the user that they can pivot to a a related data tour (i.e., the yellow and blue tours in \autoref{fig:tours-concept}). If the user decides to embark on a new tour, the new tour will be started (\gbalance). Likewise, \tool{} suggests related tours at the end of each tour. 

\subsection{Personalizing and sharing data tours}
The last option to introduce flexibility (\gbalance) is to personalize tours or create new tours. For example, slides of interest can be bookmarked in any tour using a star button. An analyst can edit any data tour by clicking the ``\textit{Edit the tour}'' button at the bottom of the data tour panel. In the editor panel, they can add new slides from \tool' library of \numfacts{} facts. Selecting interesting tags (e.g., \texttt{weight}, \texttt{outliers}, and \texttt{link}) of the slides, filters the list of slides to display only those that are relevant for completing or generating new sections with a central topic. The analyst can also change the order of slides in a given tour, remove irrelevant slides from a tour, or create entirely new tours from the fact library. Personalized tour templates can be exported in JSON format and shared with peers for reuse. Please refer to the supplementary video for additional details.

\section{Implementation Notes}
\label{sec:implementation}
\tool{} is implemented as an open-source web-application. We used D3.js~\cite{d3} for data processing and manipulation, flowmap.gl~\cite{flowmap} and deck.gl~\cite{deckgl} to render geospatial network visualizations, mapbox~\cite{mapbox} to render the underlying map, The Vistorian library~\cite{alkadi2022understanding} to create interactive node-link diagrams, and React~\cite{react} to build the interactive interface. 

\subsubsection*{\textbf{Internal Tour Specification}}
\label{sec:template-format}

In \tool{}, tours are defined by a JSON specification, making the creation of new templates straightforward and flexible for developers. A specification defines a tour's \texttt{id}, \texttt{name}, and \texttt{scope}. The \texttt{scope} describes which network elements need to be defined by the analyst at the beginning of a tour: the overall network (nothing needs to be \nobreak selected), a subgraph, or a single node. The specification then contains \texttt{sections} and their respective array of \texttt{fact}-IDs (slides). Examples of JSON specifications can be found on our website: \url{https://networknarratives.github.io/tours}.


\subsubsection*{\textbf{Tagging Facts}}\label{sec:tags}

\tool{} has a library of \numfacts{} fact templates about networks, including information about the network, subgraphs or individual nodes and links. Facts about centrality measures, topological information, attribute information, comparisons, rankings, outliers, paths, geographic information, clusters, or connectivity trends are included. A fact template is described by four attributes: 

\begin{itemize}
    \item \textbf{scope} describes what part of the network is targeted: the \textit{overall} network, a set of \textit{nodes}, or a single \textit{node}. 

    \item \textbf{link type} specifies whether a slide is suitable for networks with \textit{directed} links, networks with \textit{undirected} links, or \textit{both}. 
  
    \item \textbf{weight type} specifies whether the slide is suitable for networks with \textit{weighted} links, networks with \textit{unweighted} links, or \textit{both}. Together, the \texttt{link type} and \texttt{weight type} define what networks a slide can be applied to. 

    \item A fact can have one or multiple \textbf{tag}s (usually 2-3), which are used when selecting slides to recommend as detours. Tags are not meant to be exhaustive nor mutually exclusive. Tags for each fact were chosen by two of the paper's co-authors and currently include a total of 15 terms, such as \texttt{geography}, \texttt{nodes}, \texttt{links}, \texttt{weight}, \texttt{outliers}, \texttt{connectivity}, \texttt{statistics}, \texttt{density}, and \texttt{extrema}. 
\end{itemize}

\subsubsection*{\textbf{Fact Recommendation}}





When the user requests a detour, i.e., additional slides for a slide or section (\autoref{sec:ui_demo}), \nobreak \tool{} identifies the most relevant slides to display. First, it counts the number of the occurrences of each topic tag in each section and calculates the TF-IDF score~\cite{salton1986introduction} as a measure of the importance of each tag to each section; this approach gives a higher weight to tags that are used more often in a section than to other tags, but decreases the weight of tags that are also used frequently in other sections. 
Each tour section is then represented by a vector, in which the $i$'th element is the TF-IDF score for that section and the $i$'th keyword.
We calculate the cosine similarity between these vectors and return three randomly selected slides from the sections with the highest similarity to the section currently being expanded.
The \textit{link type} and \textit{weight type} tags are used to exclude slides that could not be applied to the network being explored.

\section{Evaluating Data Tours}
\label{sec:evaluation}
We performed two complementary user studies to understand the effectiveness of data tours in \tool{} and the extent to which we achieved the design goals listed in \autoref{sec:designgoals} (\glearn, \greduce, \gserendipidy, \grepeat, \gbalance, \gtransparency). 
One study used experts in network analysis (\autoref{sec:evaluation-experts}), whereas the other used novices (\autoref{sec:novice_evaluation}). We report on the combined results in \autoref{sec:study_results}. 

\subsection{Expert Evaluation}
\label{sec:evaluation-experts}
The expert study investigated the extent to which data tours could reduce an analyst's workload (\greduce{}, \grepeat) and whether they could provide meaningful and potentially surprising insights (\gserendipidy). Participants must have expertise in network analysis and an intrinsic interest in exploring their data to be able to compare data tours to their existing tools and workflows and to assess the usefulness of data tours. 
We invited all five analysts who were involved in the initial tour design (\markw, \chamil, \tom, \chris, \lisa) to individual exploration sessions with \tool\ by using data provided by each analyst. We added three more analysts who were not involved in the initial tour design: an analyst from academia (\pyifan) and two senior developers from an industrial communication company with 5-7 years of experience in data development (\pjiangz{}, \pweic). During individual one-hour sessions, we first demonstrated \tool{} and its main features and then provided each analyst with the tool URL, so they could run \tool{} on their own machine and explore their own data. We recorded each session, asked the analysts to think aloud and ask for help when needed, and conducted interviews to solicit additional qualitative feedback about the tours and \tool{} interface. Please refer to the supplementary material for the questions in the semi-structured interview. 


\subsection{Novice Evaluation}
\label{sec:novice_evaluation}
Our second study evaluated \tool{} with network exploration novices. It focused on understanding whether tours help learn about network concepts (\glearn) and the comprehensibility of the facts presented. 

\textbf{Conditions.} We compared two conditions: \textbf{\cguided{}} involved our \tool{} interface with all tours, the navigation panel, and full interaction within each view (pan, zoom, mouse over, and selection). We removed the tour overview panel (\autoref{fig:more_slide}) to simplify the user interface for the study and because that view is not crucial to the concept of data tours. We compared \cguided{} with the \textbf{\cfreeform} condition as a baseline because the novices would not know what to compare \cguided{} with. \cfreeform{} showed only the network visualization from \tool{}, including its interactions for exploration (pan, zoom, highlight, and selection). No tour or other navigation was available. \cfreeform{} was meant to be representative of any existing interactive free-form network visualization tool, such as Gephi~\cite{bastian2009gephi} or The Vistorian~\cite{alkadi2022understanding}.

\textbf{Participants.}
We recruited 14 (seven females and seven males) postgraduate students (N1-N14) from a local university. All participants had a background in data science but did not have any experience in network exploration. No participant had prior knowledge of \tool{} or other network exploration tools. We compensated the participants with a \pounds10 gift card for their participation. 

\textbf{Setup and data.} We ran a within-subject study with each participant experiencing both conditions. Half of the participants started with \cguided, and the other half started with \cfreeform. For each condition, the participants received a 5-min introduction to the respective interactions, followed by 10-min exploration of the network or tours. We provided two geospatial network datasets---one for each condition---with weighted and directed links. One dataset was about internal migration in Sri Lanka. It comprised 25 cities (nodes), and 600 migration routes between cities (links). The other dataset was about bicycle hires in London. It contained 786 bike stations (nodes) and 10,000 trips (links) between these stations. We chose these two datasets because they are real-world datasets of similar complexity and do not require specific domain knowledge. All participants started with the Sri Lanka migration network independently of whether they started with \cguided{} or \cfreeform{}. Each individual study lasted approximately 40 minutes. 

After each condition, we asked the participants to explore the data to gain a comprehensive understanding and describe patterns or insights that would be worth sharing with others. We asked the participants to take screenshots and note down 3-5 major findings in a 5-min time limit. We also encouraged participants to verbalize their thought process during exploration or when experiencing tours. We deliberately did not ask participants to complete a quiz because a fixed set of questions could have favored tours where facts are explicitly provided. On the other hand, tours would not have had a great chance in delivering knowledge that they did not cover. Throughout the session, the participants were allowed to seek assistance for the user interface whenever necessary and end the session early if they felt they had explored the data sufficiently. 

\textbf{Data collection.} All sessions were held online. We recorded screens, think-aloud comments, interaction logs about the number of tours and facts visited, and the time spent on each fact. At the end of the study, the participants completed a questionnaire that asked for subjective ratings about each condition on 5-point Likert scales.

\begin{figure}[t]
  \centering
\includegraphics[width=0.48\textwidth]{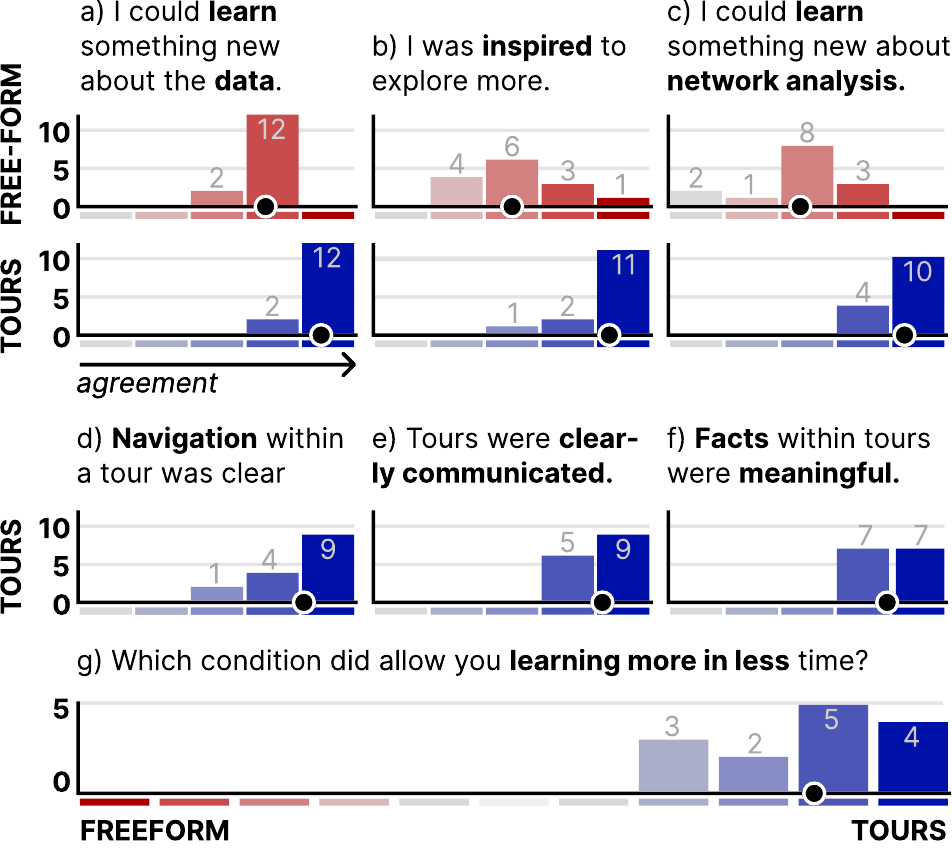}
  \caption{
  Subjective ratings from the novice study. Darker shades imply stronger agreement with the question posed. Dark circles indicate average values.
  }
  \Description{Figure 4 shows the distribution of subjective ratings. }
 \label{fig:ratings}
\end{figure}

\subsection{Study Results}
\label{sec:study_results}

This section reports the results of both studies: qualitative comments from the novices and experts, as well as subjective ratings from the novices. Generally, data tours (\cguided{}) received higher subjective ratings (\autoref{fig:ratings}) and more positive comments (\autoref{tab:table_pros_cons}) than the \cfreeform\ condition. Comments highlighted positive traits, such as \textit{orientating} and \textit{narrating}, \textit{inspiring further exploration}, \textit{leading to further discovery}, or \textit{helping learning} about analysis and exploration. With \cguided{}, participants browsed 31.1 slides (facts) on average, spending approximately 17 seconds per slide. Given a time limit of 10 minutes, the novices explored on average 2.29 tours on their own, spending 8:48 minutes with \cguided{}, whereas they spent only 5:00 minutes with \cfreeform{}. When asked \textit{``Which condition did allow for learning more in less time,''} the novices responded with an average rating of +3.71 towards \cguided{} on an 11-point scale (\cfreeform{}=-5, \cguided{}=+5, \autoref{fig:ratings}(g)). The original questionnaire showed numbers ranging from 1 (\cfreeform{}) to 11 (\cguided{}) to avoid the potentially biasing -/+ notation.). 

The novices commented that producing insights during free-form exploration (\cfreeform{}) was difficult: \says{I only get the most obvious insights; [it was] hard to further exploration and get more insights}{N10}. One novice missed guidance for a more in-depth exploration: \says{I will be attracted by the most prominent patterns, but after examining these, I feel it hard for me to do further exploration}{N14}. On the other hand, the novices found that \cfreeform{} provided more flexibility for exploration than tours and that tours could limit thinking, leading to \says{passively accepting facts}{N9}. While not entirely surprising, this observation is interesting because all free-form exploration features in  \cfreeform{} (pan, zoom, highlighting, and selection) were also present in the \cguided{} condition. We believe that this perception was due to a mindset that the study might have created because it was investigating two apparently opposing conditions. For the deployment of data tours, this situation could mean that the user interface needs to be very explicit about any free-form interaction capabilities. For example, tours should explicitly mention, explain, and encourage the use of free-form interaction with visualization. 

In the following, we focus on data tours and \tool{} in the novice and expert studies.

\begin{table}
\footnotesize
\begin{tabular}{p{0.05\textwidth}|p{0.38\columnwidth}|p{0.38\columnwidth}|}
\textbf{Cond.} & \textbf{Advantages} & \textbf{Disadvantages} \\
\hline
\hline
\cguided 
& 
Orients users with a narrative. Automatically provides facts. 
Can inspire deepened exploration. 
May lead to additional discoveries. 
Easy navigation.
Saves time. 
Helps learn about analysis. 

& Can limit thinking and feel passive.
Explanations need to be chosen carefully.
Supports a rich set of views. \\
\hline
\cfreeform 
& Provides greater flexibility for exploration. 
& Harder to obtain deep insights or spot patterns with low prominence.
Requires more time and effort to interact. 
Requires users to know where to look/have an exploration strategy. \\
\hline
\end{tabular}
\caption{Reported advantages and disadvantages for \cguided{} and \cfreeform{} across both studies.}
\Description{Table 1 compares the advantages and disadvantages of the Tours and the Free-form conditions. }
\label{tab:table_pros_cons}
\end{table}

\textbf{Tours and facts were perceived to be useful.} The novices rated tours as well communicated (\autoref{fig:ratings}(e), avg=4.71 on a 1-5 Likert scale) and found facts within the tours as meaningful (\autoref{fig:ratings}(f), avg=4.5). The analysts concurred, commenting that these facts are \says{in fact, what I would look at myself}{\chris}, and that they are \says{good for hypothesis generation}{\chris}, \says{cover[ing] a lot of common steps}{\pjiangz{}} and \says{guides what you should look at}{\chris}. For future improvements, the analysts suggested that the system could be extended by calculating additional network metrics, \says{multiple attributes at the same time}{\pweic{}}, or \says{templates that compare the layouts generated by different algorithms}{\chamil{}, \pjiangz{}}.
\tool{} was designed to make such extensions straightforward.
Likewise, the novices commented that the power of the approach grows with the number and content of the tours.

The participants highlighted the simplicity of exploring a network (\says{I prefer \tool{} because it directly guides me to see the next information}{\tom{}}) (\greduce) and that a rich set of views and information about the network is supported (\says{provides multiple perspectives of the network.}{\pyifan}), \says{The slideshow analogy is helpful [...] I can see my time saving with this}{\tom}). The novices suggested \says{I like [data tours] because the data facts are organized systematically}{N13} and \says{I can read with little effort}{N11}. Likewise, while not all facts in a given template might be of significant interest to an analyst, \tom{} expressed that facts which they would not have thought to request can nonetheless be very interesting, echoing another participant \says{Helpful to know what I really want to look more into}{\markw} (\gbalance, \gserendipidy).

\tool{} can be integrated with existing workflows, e.g., \says{to collect interesting facts and validate my observation[s]. And based on different basic facts, I still can tell different stories.}{\pyifan{}}. While \says{the existing data tours can satisfy most of my exploration needs}{\pyifan, \pjiangz}, we can easily imagine additional tours to represent bespoke exploration methods and topics. 

\textbf{Tours have educational value and inspire exploration.} The novices found that they could learn more about the data with \cguided{} (ratings in \autoref{fig:ratings}(a), \cguided=4.86, \cfreeform=3.86), found \cguided{} to teach them more about network analysis (\autoref{fig:ratings}\textit{c}, \cguided=4.71, \cfreeform=2.86), and were inspired to explore the data further (\autoref{fig:ratings}(b), \cguided=4.71, \cfreeform=3.07). \says{[Data tours] teach me how to analyze the network. Tours are like stories with different steps. I don't need to remember the key concepts. The network visualization explains well and clear[ly]}{\pyifan} (\glearn). Similarly, \tom{} suggested potential for educating students about network analysis, its concepts and methods and thought that the tours would be \says{helpful to share with the new colleagues}{\tom{}} for an introduction to network analysis. 
\pjiangz{} said \says{I like the customizing and sharing functions because new employees can use the exported tour to get familiar with the data.}{\pjiangz}
One participant highlighted the potential to generate new ideas and help develop a highly independent approach to network analysis: \says{the information can be used to answer different questions and generate new ideas (e.g., findings or assumptions) or create a new story.}{\pyifan} (\gserendipidy). 
Another participant expressed a similar thought: \says{the recommendation for querying more information around a specific topic would be useful when I have no idea about what story I could tell}{\pweic}~(\gserendipidy). 
As highlighted by a recent study~\cite{alkadi2022understanding}, this aspect is important for teaching goal and strategy development. 

\textbf{Navigation is easy.} The user interface was perceived as simple and understandable, as reflected by the average rating of 4.64 given by the novices (\autoref{fig:ratings}(d)). The analysts and novices appreciated the simplicity of retrieving and displaying facts, especially network metrics, and that relevant nodes and links were highlighted in the visualization (\says{The way of clicking the next button only is very friendly to me. I don't need other hints or reminder for what to do}{\pyifan}, \gbalance). The participants could easily navigate forwards and backwards through the sequence of slides, making it \says{easy to find back the information.}{\pyifan}. 
At the same time, the analysts suggested a set of straightforward improvements. For example, \pyifan{} suggested \says{more instruction and tooltips}{\pyifan} and \tom{} recommended that each template could have a ``\textit{very quick}'' preview, with the option to drill down and view additional details on demand. We found this observation very interesting and adapted our design accordingly (\autoref{sec:ui_demo}). Interestingly, N12 suggested providing further guidance on which tours to choose, especially when and why to pivot to a new tour.

\textbf{Widened use cases.} We discussed a range of potential applications for our approach in addition to personal exploratory analysis. \chris{} said they would use the tool to make data and qualitative exploration available to peer researchers (\says{It’s frustrating because none of the publications provide interactions for the networks. Sharing data tours is valuable to play with by the others}{\chris}). \pjiangz{} and \pweic{} mentioned that ``\textit{the customized series of sequential fact is useful for reporting and presentation}'' because they felt the interactive visualization was more vivid than the slides typically used for communication and demonstration within or outside their teams.
\tom{} suggested potential for engaging with the public by using touch screens in museums, and is currently exploring this scenario for their own research. Finally, \lisa{} suggested using these tours to facilitate communication and discussions with policymakers. 

\section{Discussion}
\label{sec:discussion}

Our study results provide strong evidence for the benefits of data tours and show that our design successfully supports our initial goals (G1-G6). The main findings can be summarized as follows:
\begin{enumerate}[noitemsep,leftmargin=*]
    \item \textbf{Tours are an extensible concept.} Our current tours showed what our analysts were interested in. Yet, the sets of facts that can be shown in data tours are potentially very large, and the tour's power grows with the number of facts that they include. New facts can easily be added to our framework, which currently contains \numfacts{} individual facts. 
    \item \textbf{Data tours are complementary to free-form tools.} The novices acknowledged that they found different insights under each condition (\cguided~or \cfreeform). They commented that data tours provide sufficient insights, but using them could be a passive activity. Thus, both conditions have their unique values to the users. 
    \item \textbf{Data tours are a means to accelerate analysis and exploration} and reduce manual labor (\greduce). Especially when dealing with numerous networks, the analysts are required to have consistency in exploration and analysis, as well as to explore these networks quickly (\grepeat). 
    \item \textbf{Quick overview can prevent analysts from getting lost} in too many options and help them keep track of their previous exploration history, e.g., using a specialized tool~\cite{heer2008graphical}. For example, the analyst can overview a network through different tours, star interesting slides, and follow up on these slides in a second iteration. 
    \item \textbf{Sequential tours support novice analysts} who are getting started with network visualization and learning about analysis methods and concepts (\glearn). They can be used by novices without previous knowledge about networks or specific network concepts or goals, who would otherwise struggle with a very open and free-form approach to network exploration~\cite{alkadi2022understanding}. Tours could help onboard and familiarize novice analysts with specific analysis routines, and even help engage and communicate networks and analysis to a broad audience.
    \item \textbf{Tours provide a serendipitous element} to exploration (\gserendipidy). Comments from the user studies suggested that tours could also provide information quickly while keeping the door open for exploration. 
\end{enumerate}


\subsubsection*{\textbf{Limitations.}}
Evaluating exploration is an intrinsically difficult problem, given its open and context-sensitive nature. Our expert study aimed to account for this difficulty, while our novice study was limited in this regard. Although we attempted to choose datasets that might have been of interest to participants, finding datasets that are of equal interest to any participant in such a study while assuring that they are of equal complexity is difficult. We deliberately did not ask for specific insights (e.g., the number of clusters) to maintain the open nature of free-form exploration in \cfreeform{} as a particular trait of that condition. We also acknowledge that data tours and free-form exploration are complementary and that future studies should compare data tours as human-designed and goal-oriented devices with pure recommender engines, such as DataShot~\cite{wang_datashot_2020}. However, at the time of our research, none of the existing recommender engines supported network analysis. Such a comparison could help assess to what extent these recommender systems are as transparent (\gtransparency) as our data tours.

Although our study had only 10 tours available, we believe that they are representative of common network exploration tasks. We are now deploying \tool{} as an add-on to The Vistorian~\cite{alkadi2022understanding} to understand data tours being used over a long period of time. We also aim to deploy \tool{} to students of network analysis. However, such an effort would require further discussions with educators on how to best use \tool{} in a curriculum.


\subsubsection*{\textbf{Handling complex types of networks and analyses}}
Currently, \tool{} can handle network data that are weighted or unweighted, geospatial or nongeospatial, and temporal. Although it does currently not support networks with multiple node and link types, adding respective facts to our slides would be trivial. Fully supporting temporal networks, however, appear to be a major challenge with respect to the tasks (facts) supported (e.g.,~\cite{Ahn2011}) and to the performance of meaningful temporal analysis to obtain facts~\cite{xie2020interactive}. The challenge is that complex types of networks also require complex analyses, for example, understanding and analyzing changes in dynamic networks~\cite{xie2020interactive}. Although task taxonomies~\cite{ahn2013task} can be a starting point for design, we also need to better understand further strategies and workflows that analysts employ in practice to explore these complex types of networks. 


\subsubsection*{\textbf{Additional visual representations and presenting formats}}
Expanding the scope of \tool{} to include complex types of networks and a richer collection of facts will require additional types of visualizations. This extension may include network visualizations~\cite{Beck2016,schoettler2021visualizing} or visualizations to summarize graph metrics, such as line graphs, bar charts, or ranked lists (e.g.,~\cite{SYF}). 
Going beyond well-known visualization techniques and including a more diverse set of visual representations into tours will, in turn, require explanations of these techniques and their respective visual encodings (e.g.,~\cite{wang2020cheat}). We are also investigating other ways of presenting networks, such as data comics~\cite{bach2016telling} or videos~\cite{amini_understanding_2015} and the extent to which these formats foster engagement and understanding. 

\subsubsection*{\textbf{Overcoming challenges to automation}}
Automation can help extract a wide range of facts, such as network motifs, which we excluded from our current version for reasons of computational complexity. However, open questions as to the extent an algorithm can identify qualitative \textit{insights} into networks remain. For example, we can calculate network metrics and topological features (motifs, paths, communities, and bridge nodes). However, how do we teach a machine to look for more subtle or complex patterns? Exploratory data analysis is powerful \textit{because} it is done by humans.


Although machines can support this process by suggesting facts about data, automatically generating insights that incorporate users' domain knowledge and mental models will be difficult. One promising direction could be to begin by focusing on developing data tours for specific types of networks, such as biological pathways or social networks. These networks would have specific terminology and involve similar tasks, to which tour templates could be tailored. 

Eventually, we imagine data tours being adjusted on-the-fly on the basis of the data being explored. For example, if a network has a ``complex'' community structure, a tour might provide additional information about that aspect. Likewise, if a network has many distinct communities, then a tour might report a few facts about each community. We believe that our linear, yet human-scripted, data tours are a first step toward complex tours that contain branches, loops, and other structural constructs. Our approach could inspire recommender systems that include further guidance and goal-oriented exploration. While understanding an analyst's analysis goals from interactions with a system is a well-known problem, data tours could propose directions then adapt to both data, and user interaction, while remaining direction.




\subsubsection*{\textbf{Keeping data tours concise}}
Given that the collection of data facts and tour templates may continue to grow, keeping data tours concise will require active effort. Short and concise data tours were identified as an important criteria during our interviews. 
However, data tours need to contain sufficient facts to provide meaningful insights into networks. 

We can address this situation in part by extending our data tours through appending additional related facts to sections or slides, as described in \autoref{sec:ui_demo}.
Additional approaches that could help keep data tours concise are available. One option is to split tours and create extended versions of the existing tours, which are designed \textit{a-priori}. This approach might support expert analysts in performing complex network analyses. An alternative is to create direct sequels that pick up where a previous tour left off, but otherwise follow their own logic. Another option could be to use \textit{starred} slides to create a user profile, akin to other existing systems (e.g., ~\cite{shi_calliope_2021}). We believe that adding this technique to \tool{} could be straightforward because it is conceptually independent from the dataset. Eventually, we could assign each fact in a tour a ``priority'' score and shorten or extend the tour on a user's demand (which could be expressed explicitly by using a simple slider, or inferred from user's interaction history by using a machine learning approach).


\subsubsection*{\textbf{Support personalization, sharing, and storytelling}}

The growing number and size of possible data tours and facts suggests the potential for the personalization of tours and use of bespoke storytelling approaches to maintain clarity and improve communication. The creation of personal templates in \tool{} is currently supported through \textit{starring} slides and editing tours. In the future, tour templates could be shared, commented on, modified by others, and reshared like software in the open-source community or plans for physical models in the maker community~\cite{makerbot2013thingiverse}. 

While \tool{} is explicitly not designed for storytelling, data tours provide considerable potential for storytelling and presenting insights. Manually created data tours could be described as stories, especially if made by the network analyst themselves. However, it may require proper narrative language---in contrast to the currently rather factual explanations---that could include questioning, sign posting, analogies, metaphors, and other elements of narration to automate engaging storytelling~\cite{bach2018narrative}. Some of these extensions are a straightforward matter of scripting data tours in the form of personal stories and using story templates as suggested by \tom. We believe in the existence of an open space for future research to support storytelling about network data beyond current approaches, such as static data comics for networks~\cite{bach2016telling,kim2019datatoon}, or interactive comics that combine explanation and exploration~\cite{wang2021interactive,kang2021toonnote},  slideshows~\cite{satyanarayan2014authoring}, and data videos~\cite{amini2016authoring} for visualization. However, as illustrated by the large variety of narrative structures proposed in the literature, no universal recipe for creating good narratives exists.
Storytelling is a very human activity that is inherently difficult to automate.
\section{Conclusion}
\label{sec:conclusion}

In this paper, we explored semi-automated data tours to aid network exploration and help learn concepts of network analysis. Our work is the first to explore data tours in the context of networks. Rather than implementing a full-fledged recommender engine, we opted to create goal-oriented tours inspired by real-world analysis practices. We see data tours as complementary to recommender approaches, as well as free-form exploration. At the same time, \tool{} attempts to strike a balance between statically defined tours, data-driven recommender systems, and open-ended free-form exploration by including semi-automated techniques to retrieve related facts on user demand and allow for basic forms of personalization and exploration. We created an initial set of 10 data tours, and we imagine that additional data tours will be shared, modified, and reshared among analysts, potentially creating a global repository of tours. Feedback from expert analysts and novices suggests great potential for saving time and manual labor while helping to orient users. 
We believe that there is a promising future for future tools that can automatically identify and visually communicate insights by drawing on storytelling techniques. 

\begin{acks}
The authors would like to thank the experts and participants for their help in the project, as well as the anonymous reviewers for their valuable comments. This work is partially supported by Hong Kong RGC GRF Grant (No. 16210321), a grant from MSRA, EPSRC (Project EP/V010662/1, EP/T517884/1), and NSFC (No. 62202105). 
\end{acks}

\balance

\bibliographystyle{ACM-Reference-Format}
\bibliography{bib/long_dois}



\end{document}